\documentstyle[twocolumn,prl,aps,epsf]{revtex}

\begin{document}

\draft

\twocolumn[\hsize\textwidth\columnwidth\hsize\csname
@twocolumnfalse\endcsname

\title{van der Waals interaction in nanotube bundles: consequences
  on vibrational modes}

\author{Luc Henrard$^{1,2}$, E.~Hern\'andez$^3$, Patrick Bernier$^2$ and Angel Rubio$^4$}
 \address{$^1$Laboratoire de Physique du Solide,
  Facult\'es Universitaires Notre-Dame de la Paix. 5000 Namur. Belgium\\
  $^2$ Groupe de Dynamique des Phases Condense\'es. Universit\'e
    de Montpellier II. 34095 Montpellier. France\\
  $^3$ School of Chemistry, Physics and Environmental Science, 
  University of Sussex, Brighton BN1 9QJ, England UK\\
  $^4$Departemento de F\'\i sica
  Te\'orica, Universidad de Valladolid, E-47011 Valladolid, Spain}

\date{\today}
\maketitle
\begin{abstract}
  We have developed a pair-potential approach for the evaluation of
  van der Waals interaction between carbon nanotubes in bundles.
  Starting from a continuum model, we show that the intertube modes 
  range from $5 cm^{-1}$ to $60 cm^{-1}$. Using a non-orthogonal tight-binding
  approximation for describing the covalent intra-tube bonding in
  addition, we confirme a slight chiral dependance of the breathing
  mode frequency and we found that this breathing mode
  frequency increase by $\sim$ 10 $\%$ if the nanotube lie inside a bundle 
  as compared to the isolated tube.
\end{abstract}
]

Research in carbon nanotubes is now a very active field both because
of their fascinating cylindrical structure and potential applications
(see~\cite{dresselhaus_96,terrones_99}). On the characterization side,
the direct measurement of atomic structure is now possible by means of
scanning tunneling microscopy (STM) experiments~\cite{STM}. This
technique gives information on both the local atomic structure as well
as electronic properties. However it is too time-consuming for routine
sample characterization. Thus one has to resort to more macroscopic
techniques, which are able to handle the whole as-grown sample.  One
of the most popular alternative experimental tools is the measurement
of the vibrational spectrum by Raman spectroscopy (see e.g.
\cite{raman}).  Since the suggestion was made \cite{raman} that the
vibrational frequency of the fully symmetric $A_g$ breathing mode (BM)
of an isolated nanotube could be used to determine the nanotube
diameter, given its marked frequency dependence on the diameter of the
tube, this idea has been widely used as a characterization tool.  But,
it should also be noted that a small chiral dependance of BM frequency
was recently found by ab-initio calculations
\cite{kurti_98,sanchez_99}.

However for developing useful technological applications, it is of
primary importance to refine the characterization techniques of
nanotube samples, taking into account the local environment in which
the nanotubes find themselves. Surprisingly, theoretical studies of
the resonance frequency have largely focused on isolated tubes, and no
such studies have been reported for bundles of tubes, which is the
form in which the majority of the nanotubes are to be found in
experimental samples. The aim of the present communication is to show
how the tube-tube interactions can be taken into account for the
characterization of the BM. Due to its radial character, the BM is
likely to be the most influenced mode by the nanotube packing and
therefore it is neccesary to quantify this effect.  But before we
address this question, we will present an evaluation of the inter-tube
vibrational modes, which could also allow the characterization of
samples by inelastic neutron scattering experiments.
  
The theoretical model we have developed to deal with the tube-packing
effects in the vibrational spectra combines a description of the
covalent bonds inside each nanotube via a non-orthogonal tight-binding
parametrization~\cite{tb_note}, which has proved to work very well for
the structural and mechanical properties of carbon
nanotubes~\cite{hernandez_98}, with a pair-potential approach to deal
with the van der Waals interaction between carbon nanotubes. The
long-range dispersion interaction is described by a carbon-carbon
Lennard-Jones potential $V_{cc}$ given by $
V_{cc}=-\frac{C_6}{d^6}+\frac{C_{12}}{d^{12}} \; ,$ where $d$ is the
carbon-carbon distance and $C_6$ and $C_{12}$ are constants fitted to
reproduce the structural properties of graphite~\cite{constants}.
Despite its intrinsic simplicity, the present approach is widely used
to simulate van der Waals interactions between molecules (see
\cite{israevlachvili_92}). In all the calculations described below we
consider infinitely long tubes. Let us also note that, although
ab-initio descriptions are in general more accurate and reliable, they
are very computationally demanding and are limited in the size of the
system~\cite{sanchez_99,rubio_99}.

First, to study the intertube modes, we simplify further the model and
we consider carbon nanotubes as continuous cylindrical surfaces of
density $\sigma$. In this case we can compute analytically the
potential felt by a carbon atom situated at a distance $a$ from the
center of the nanotube of radii $R$. The potential reads ($a>R$):
\begin{eqnarray}
V_{cT}(a)=&& \frac{3}{4}\pi R \sigma \left\{ \frac{C_6}{a^5}
F\left[5/2,5/2,1,(R/a)^2\right] \right. \nonumber \\
 && \left. +\frac{21
C_{12}}{32 a^{11}}F\left[11/2,11/2,1,(R/a)^2\right]\right\} \; ,
\end{eqnarray}
where $F$ is the hypergeometric function \cite{abra}. This model
cannot distinguish between different layer-stackings in multilayer
compounds, nor can it take account of the tube chiralities when
applied to bundles of nanotubes.  We feel that this is a good
approximation as the registry property in nanotube bundles is
frustrated by the geometry of the bundle. Our model thus can be
considered to represent an average over the different stacking
possibilities in a bundle of tubes of similar diameter having
chiralities compatible with that diameter.  In this sense we expect
the continuous model potential to be able to give a realistic
description of the intertube modes.  This approach is similar to the
one developed by Girifalco for the $C_{60}$ fullerene
\cite{girifalco_92}.  Within this simplified model the van der Waals
interaction energy per unit length between two carbon nanotubes,
$V_{TT}$, follows immediatly by numerical integration of $V_{cT}(a)$
over the surface of the second nanotube. Now, the total energy of a
bundle with $N$ tubes, $V_B$, is obtained by summing $V_{TT}$ over all
tube-pairs. Before the computation of the vibrational spectra, the
total energy for a given bundle is minimized with respect to the tube
positions~\cite{relax}. Then, the intertube vibrational modes
(eigenvalues and eigenvectors) are deduced from the diagonalisation of
the dynamical matrix constructed from a numerical evaluation of the
second derivatives of the total energy with respect to the tube center
coordinates~\cite{comment_c60}.  As a practical remark, we have built
the finite-size bundles enforcing the 6-fold symmetry in the section
of the bundle, which has one nanotube at its center (the number of
tubes is $N=6i+1$ with $i$ being the number of hexagonal-shells in the
bundle). In this case, all the inter-tube vibrational spectra can be
sorted according to the $C_{6v}$ point group. In particular we find
$3i$ Raman active modes and $2i$ infrared (IR) active modes.

\begin{figure}
\epsfxsize=3 in
\epsfbox{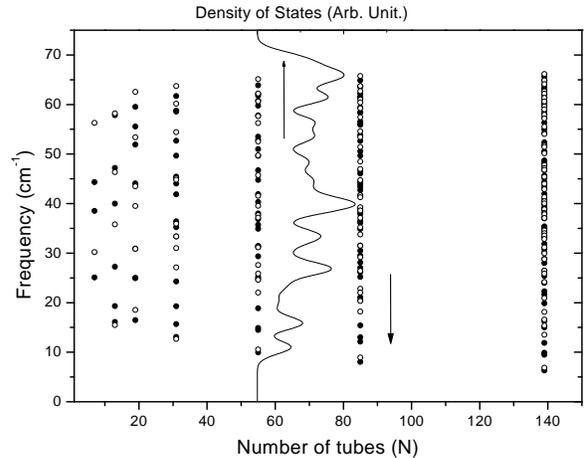}
\caption{Intertube mode frequencies of $C_{6v}$ bundle of infinitely long
  nanotubes of radius $R$=6.8\AA~as a function of the number of
  tubes. Only Raman active ($A_1$ and $E_2$) modes and IR modes
  ($E_1$) are represented by fill/open circles . The total vibrational
  density of states, using a gaussian broadening of $0.8 cm^{-1}$, for
  a bundle with 55 tubes is also given.
}
\label{inter}
\end{figure}

In Fig.~\ref{inter} we show the Raman and IR active modes of a finite
size bundle of (10,10) ($R$=6.8\AA) armchair nanotubes as a function
of the number of tubes $N$. Intertube modes are found between $5
cm^{-1}$ to $60 cm^{-1}$.  To the best of our knowledge, no experimental
evidence of the observation of Raman active modes in this energy range
have been reported. The origin of these lack of observational
evidences is probably a low scattering cross section and difficulties
to deconvoluate the elastic-scattering peak.  Inelastic neutron
scattering experiments should also be able to probe such modes even if
the difficulty of rigid displacement of long, massive cylinders could
reduce the excitation probability.
However such experiments require a large quantity of highly purified nanotube
powder and at present no data is available for the inter-tube
vibration energy range \cite{rols_neutron}. For comparison between our
simulation and future experimental work, we show by the solid curves
in figure \ref{inter} the total vibrational density of states of a
bundle consisting of 55 tubes.  Note also that the squashing mode of
isolated tubes and the libration mode~\cite{kwon_98} are predicted at
about $16 cm^{-1}$ \cite{libration} and are likely to interfere with intertube
modes.  This last mode is expected to play a role in temperature
dependant conductivity of nanotube bundles since it can induce
temperature dependent hopping conductivity. The intertubes mode
described here could have similar behaviour.

After presenting the new data on intertube vibrational modes, we look
at the influence of packing in the BM. The calculations were performed
in a frozen phonon approach using the hybrid tight-binding plus
continuous Lennard-Jones potential described above. We use a
conjugate-gradient relaxation scheme to determine the geometry of the
isolated tube before making the full-relaxation of the tube-bundle.
After relaxation, we applied the frozen phonon approach to evaluate BM
frequencies by computing the total energy (intra- and inter- tube) for
a 0.1 $\%$ change of the tube radius.  One of the limitation of the
frozen phonon calculation for the BM in bundles is the fact that
the pure radial mode is no more an eigenmode of the bundle system since
packing breaks the symmetry of isolated tubes. This will lead, in
addition to the frequency shift, to a degeneracy lift. These
effect were study for $C_{60}$ in \cite{yu_94}. Here we do not take
this symmetry lowering in consideration but we believe that frozen
phonon approach catches the essential feature on the packing effect on
BM modes and gives a first good evaluation of the resonance frequency
increase that will be usefull for Raman spectroscopy analysis of
nanotube sample.

\setbox1=\hbox{\cite{steph_99}}
\setbox2=\hbox{\cite{sanchez_99}}
\setbox3=\hbox{\cite{kurti_98}}
\begin{figure}
\epsfxsize=3 in
\epsfbox{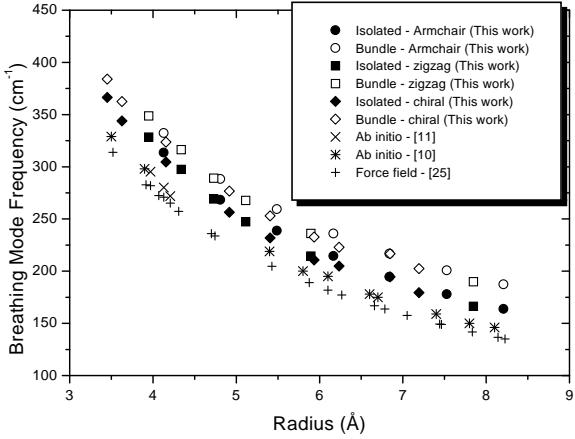}
\caption{Breathing mode ($A_1$) frequencies as a function of the tube
  radius. This work : Circles are for armchair tubes, squares for
  zig-zag and diamond for chiral tube. Solid symbols are for isolated
  tubes whitin tight-binding approach and open symbols are for tubes
  in a bundle (see text). $+$ are estimated from force field approach
  \box1, $\times$~\box2~and $\star$~\box3~come from ab-initio calculations.
}
\label{BM}
\end{figure}

In Fig.~\ref{BM} and Table~\ref{table} we present the results for the
BM modes of isolated tubes of different radii and chiralities as well
as the values for the corresponding infinite bundles.  As far as
isolated tube concerns, we see that armchair $(n,n)$ and zigzag
$(n,0)$ tubes do not follow exactly the same scaling law with
diameter. We reproduce the known $1/R$ scaling of the BM frequency
($\nu$), in particular the fit of our data to $\nu=C/R$ give the
following values: $C=$1307$\mbox{cm}^{-1}$~\AA\ for armchair tubes and
$C$=1282$\mbox{cm}^{-1}$~\AA\ for zigzag tubes.  Chiral tubes BM mode
lie between armchair and zigzag lines. Recent ab-initio results
\cite{kurti_98} showed $C$=1180$\mbox{cm}^{-1}$~\AA\ and $C$=1160$
\mbox{cm}^{-1}$~\AA\ for armchair and zigzag tubes, respectively and
force field data are reproduced by $C$=1111$\mbox{cm}^{-1}$~\AA\ 
\cite{steph_99} or $C$=1147$\mbox{cm}^{-1}$~\AA \cite{popov_99}
irrespective of tube chirality.  Considering that tight-binding
methods usually overestimate vibrational frequencies by 5-10 $\%$
\cite{porezag_95,BM_c60}, the results presented above are not
surprizing. We emphasize that our tight-binding calculation reproduce
well the slight difference between armchair and chiral tubes found by
ab-initio methods~\cite{sanchez_99}. The reason for this success is
connected with the non-orthogonality of our tight-binding method that
correctly describe the re-hybridization of carbon-carbon bonds when
the nanotube breathe.  Moreover, the low computational cost of the TB
approach compared to DFT allowed us to perform a more systematic
study of chiral and achiral nanotubes (see Table \ref{table}).

We can now turn to the case of nanotube bundles.
As expected, Fig.\ref{BM} and Table \ref{table} show a clear
increase in the BM frequencies when tubes are packed into
bundles. The relative increase goes from 5 $\%$ for a tube of radius
$R$=3.5\AA~to 15 $\%$ for $R$=8\AA. For $R$=6.8\AA~
(that are revelant considering actual produced nanotubes), the
frequency shift is evaluated to be of the order of 10 $\%$. Very
recently, Venkateswaran et al.\cite{venkateswaran_99}, using a similar
model to study the pressure dependance of nanotube bundles Raman
modes, found a $8\%$ increase of the BM of (9,9) tube when they are packed
compare to isolated tube. This is consistent with the present result
and the difference (we found $10\%$ increase for a (9,9) tube) is
likely to come from a sligth difference in parameters used both for
the tight binding and for the pair van der Waals potential. 

Focussing on consequences on the interpretation of Raman spectra, the
effect of neighboring tubes on BM frequencies can lead to
misinterpretation of Raman results and to a 10 $\%$ error in radius
determination. Even if tight-binding calculations do not give exact
numbers for vibrationnal energies, our results are inclined to
conclude that the experimental Raman peak at $180 cm^{-1}$ is to be
associated with tubes of diameter larger than the $(10,10)$ tube
diameter. Indeed, simulation for single-wall nanotubes predicted a BM
for a (10,10) tube between $163 cm^{-1}$ (force field \cite{steph_99})
and $194 cm^{-1}$ (this work) or at $178 cm^{-1}$ for the ab-initio
evaluation in \cite{kurti_98}. If we consider those results as 10 $\%$
under-evaluation of the BM of tube in a bundle, we get frequencies
higher than $180 cm^{-1}$ (except for the less sophisticated force
field model) and the key frequency of $180 cm^{-1}$ is associated with
tube with diameter larger than $6.8 \AA$. This conclusion is
consistent with X-rays \cite{journet_97,thess_96} and neutron
\cite{rols_99} diffraction if the tube-tube distance is taken to be
$3.2 \AA$ (and not the graphite interlayer distance $3.35\AA$) since
the diameter polydispersity required for diffraction spectra fit
\cite{rols_99} emplies the presence of tube larger than the (10,10)
tube in bundles. Moreover, electron diffraction
\cite{henrard_99} on single nanotube rope lead to the conclusion that
$6.8 \AA < R < 7.5 \AA$ and then to a mean radius larger than the the
$(10,10)$ radius.

As the BM concerns, we have only considered bundles made of a infinite
number of nanotubes and the study of finite bundles will obviously
lead to a gradual change of the BM vibration frequency going from a
frequency close to the bulk-value for
the central tubes (surrounded by 6 tubes) to a frequency close to 
the isolated-tube value corresponding to tubes at the bundle-surface.
This behavior is expected from the  short-range tube-tube interaction
(at the scale of tube diameter). Note that
STM \cite{STM} and electron diffraction
\cite{henrard_99} experimental studies of carbon nanotubes
concluded that all tube chiralities
can be found in samples with a very narrow diameter distribution.
This will not change drastically our conclusions, even more, if tube
chirality are randomly distributed whithin the rope, no registry
is possible between adjacent tubes and the continuous model is well
justified. Furthermore, we have checked that an increase/decrease of the
tube diameter within the rope leads to a reduction/increase of 
the BM vibrational frequency of neighbouring tubes.
Then, both the surface effect and 'defect'
(larger or smaller tubes) will then lead to a broadening of the
experimental Raman spectra as observed experimentaly.

\begin{table}
\caption{Breathing Mode (BM) frequencies for various carbon nanotubes.
 The first column
  defined the tube, the second one is the radius of the relaxed
  structure. The frequencies of the isolated tubes and tubes in
  bundle are given in the next two columns. The last column is the
  relative increase of the BM frequency when bundles are packed.
\label{table}}
\begin{tabular}{ccccc}
$(n,m)$&$R (\AA)$&$\nu_{isol}(cm^{-1})$&$\nu_{bundle}(cm^{-1})$&shift($\%$)\\
\tableline
$(6,4)$&$3.45$&$366$&$384$&$4.8$\\
$(8,2)$&$3.63$&$344$&$362$&$5.3$\\
$(7,4)$&$3.81$&$313$&$330$&$5.2$\\
$(10,0)$&$3.91$&$328$&$349$&$6.2$\\
$(6,6)$&$4.07$&$313$&$332$&$6.0$\\
$(10,1)$&$4.16$&$304$&$323$&$6.3$\\
$(11,0)$&$4.34$&$297$&$316$&$6.4$\\
$(12,0)$&$4.70$&$269$&$289$&$7.4$\\
$(7,7)$&$4.81$&$268$&$288$&$7.4$\\
$(10,4)$&$4.92$&$256$&$276$&$7.9$\\
$(13,0)$&$5.11$&$247$&$268$&$8.3$\\
$(12,3)$&$5.41$&$232$&$253$&$9.1$\\
$(8,8)$&$5.49$&$239$&$259$&$8.6$\\
$(15,0)$&$5.90$&$214$&$236$&$10.1$\\
$(14,2)$&$5.93$&$211$&$233$&$10.4$\\
$(9,9)$&$6.17$&$214$&$236$&$10.1$\\
$(12,6)$&$6.23$&$205$&$227$&$10.7$\\
$(10,10)$&$6.85$&$195$&$217$&$11.4$\\
$(16,4)$&$7.19$&$179$&$202$&$12.8$\\
$(11,11)$&$7.53$&$178$&$201$&$12.9$\\
$(20,0)$&$7.84$&$166$&$190$&$14.2$\\
$(12,12)$&$8.21$&$164$&$187$&$14.4$\\

\end{tabular}
\end{table}

In conclusion, we have presented the first study of the
inter-tube vibrational modes (both Raman and IR active)
in bundles of single walled nanotubes and proposed
neutron inelastic scattering as a experimental test of the validity of
our empirical model. We have also showed the first
computational evidence of the packing influence on the BM of nanotubes
and drawn conclusions on the interpretation of the
experimental Raman spectra and how to extract useful experimental
information about the nanotube structure and diameter.

{\it Acknowledgments:} We have benefited from fruitful discussions
with P.Senet, E.Anglaret, S.Rols, J.L.Sauvajol, A.Loiseau and
Ph.Lambin. This work was supported by the TMR contract
NAMITECH(ERBFMRX-CT96-0067(DG12-MIHT)), the Belgian Program(PAI/UAP
4/10) and JCyL(VA28/99). L.H. is supported by the Belgian Fund for
Scientific research (FNRS).

\bibliographystyle{prsty}


\begin{thebibliography}{10}
\bibitem{tb_note} A non-orthogonal tight-binding $\sigma+\pi$ scheme
  was used using 11 irreducible points over the one-dimensional
  Brillouin-zone of the isolated tube. The hopping integrals and the
  hamiltonian and overlap matrices are tabulated as a function of the
  internuclear distance on the basis of first principle DFT
  calculations and only one and two center contributions to the
  hamiltonian matrix were retained (see a detailed description in
  \cite{porezag_95,goringe_97}).
\bibitem{relax}
For an infinite 2D-packing of $(10,10)$ tubes with radii of $R$=6.8~\AA~the
  relaxed intertube distance is found to be $3.14~\AA$. Note that the value for
  the interlayer distance between two flat carbon sheets within the continuous
  model is $3.27 \AA$ and for an infinite layer stacking (graphite)
  is $3.23 \AA$. This distance correspond to the actual interlayer distance in
  graphite $3.34~\AA$. The $\sim$0.1\AA difference is a measure of the
  graphene layer thickness. Then, we have confirmed the assumption that
  curvature reduces the intertube distance in the bundle as compare to graphite
  by more than $0.1~\AA$.
\bibitem{comment_c60} We note that for the calculation of phonons in
  $C_{60}$ solid, the Lennard-Jones carbon-carbon intermolecular
  potential has been succesfully used together with a bond charge
  model \cite{yu_94}. This last part of the potential was added
  empirically to reproduced the relative orientation of $C_{60}$
  molecules in the solid. In the case on nanotube-bundles there is no
  evidence of such orientation and we do not include such bond charge
  model in our calculation.
\bibitem{libration} This libration mode do not come out the present
  simulations because of the continuum character of the tube-tube
  potential we have used.
\bibitem{BM_c60}
The same scheme for $C_{60}$ gives a BM frequency at $550 cm^{-1}$
where experimental value is $497 cm^{-1}$.
\end{thebibliography}




\end{document}